\documentclass[preprint,12pt]{elsarticle}
\usepackage{amssymb}
\usepackage{amsmath}
\usepackage{graphicx}
\usepackage{hyperref}
\usepackage{xcolor}
\usepackage{orcidlink}

\biboptions{sort&compress}
\newcounter{bla}

\journal{Computer Physics Communications}

\begin{document}

\begin{frontmatter}

\title{Numerically Optimizing Shortcuts to Adiabaticity: A Hybrid Control Strategy}

\author[a,b]{Bo Xing\corref{author1}~\orcidlink{0000-0002-9651-5505}}
\author[c]{Jes\'us G. Parejo}
\author[c,d]{Sof\'ia Mart\'inez-Garaot~\orcidlink{0000-0002-6916-3858}}
\author[a,e,f]{Paola Cappellaro~\orcidlink{0000-0003-3207-594X}}
\author[a,d,g]{Mikel Palmero\corref{author2}~\orcidlink{0000-0001-9222-5298}}

\cortext[author1] {boxing92@mit.edu}
\cortext[author2] {mikel.palmero@ehu.eus}
\address[a]{Research Laboratory of Electronics, Massachusetts Institute of Technology, Cambridge, Massachusetts 02139, USA}
\address[b]{Quantum Innovation Centre, Agency for Science, Technology and Research, Singapore 138634, Singapore}
\address[c]{Department of Physical Chemistry, University of the Basque Country UPV/EHU, Apartado 644, Bilbao 48080, Spain}
\address[d]{EHU Quantum Center, University of the Basque Country UPV/EHU, 48940 Leioa, Spain}
\address[e]{Department of Nuclear Science and Engineering, Massachusetts Institute of Technology, Cambridge, Massachusetts 02139, USA}
\address[f]{Department of Physics, Massachusetts Institute of Technology, Cambridge, Massachusetts 02139, USA}
\address[g]{Department of Applied Physics, University of the Basque Country UPV/EHU, Bilbao 48013, Spain}

\begin{abstract}
Achieving fast, excitation-free quantum control is a vital challenge in modern quantum technologies. In many cases, shortcuts to adiabaticity enable fast adiabatic-like protocols, yet determining control parameters that satisfy practical constraints is often challenging in complex systems. Here, we combine an analytical shortcut to adiabaticity approach with several numerical optimization methods to boost the performance of the protocol. As a proof-of-principle for this hybrid approach, we study a particularly intricate control problem, the separation of two trapped ions. We show that this analytical-numerical approach, along with the physical insight gained through the variety of suboptimal solutions, leads to the exploration of new solutions in a complex landscape that yield improvements of up to 3 orders of magnitude. Moreover, this improvement comes with no additional cost from an experimental point of view.
\end{abstract}

\end{frontmatter}

\section{Introduction}
Shortcut-to-adiabaticity (STA) methods provide fast routes for quantum control that reproduce the outcome of slow adiabatic evolutions in shorter evolution times~\cite{Torrontegui2013, GueryOdelin2019}, drastically reducing its exposure to decoherence.
They comprise a set of mathematical techniques that engineer control protocols to reach the same final state as an adiabatic process, without necessarily following the adiabatic trajectory, thereby reducing the required driving time~\cite{Chen2010, Demirplak2003, Masuda2009, MartnezGaraot2015}.
Invariant-based inverse engineering \cite{GueryOdelin2019, Chen2010}, in particular, constitutes a major framework for designing STA protocols by imposing boundary conditions on auxiliary functions. 
In realistic implementations, however, these constructions leave a set of free parameters that must be determined numerically under additional physical constraints. 
The resulting optimization problems are typically nonlinear, sensitive to model approximations, and characterized by unbounded parameter spaces. Narrow valleys, disconnected local minima, and regions of unphysical solutions may emerge, making convergence strongly dependent on the choice of numerical method. 
Understanding how different optimizers explore this free-parameter landscape is therefore essential for developing robust STA protocols for complex quantum control problems.

A concrete and experimentally relevant example arises in the manipulation of trapped ions, which is central to the Quantum Charge-Coupled Device (QCCD) architecture proposed by Wineland and co-workers~\cite{Kielpinski2002}. 
Imperfect control and prolonged interaction with the environment introduce decoherence in the quantum states of interest, degrading the information they encode and compromising the fidelity of the operation. 
The incorporation of STA techniques has enabled more accurate and efficient control in many tasks, including but not limited to shuttling, trap expansions, ion rotations, and ion separation~\cite{Palmero2014, Palmero2015b, Palmero2015, Palmero2017, Tobalina2021}. 
Yet, despite the widespread use of numerical optimization in STA, the structure of the associated optimization landscapes and their influence on algorithmic convergence remain poorly understood, particularly in protocols that involve strong nonlinear effects.

Among these, ion separation represents the most demanding control problems~\cite{Home2006} due to the presence of trap deformations that break harmonicity. 
Several experiments separating two trapped ions~\cite{Rowe2002, Bowler2012, Ruster2014, Pino_2021, Wan2019, Lancellotti2024} have highlighted these difficulties, and theoretical proposals have sought to mitigate the associated technical challenges~\cite{Palmero2015, Sutherland2021, Simsek2021, Guglielmo2024}. 
However, mathematical complications arising from the Coulomb interaction between the ions severely restrict the effectiveness of analytical treatments~\cite{Palmero2015, Lizuain2017}. 
Once anharmonic effects are included, simultaneously satisfying all boundary conditions while minimizing residual excitation becomes increasingly challenging. 
In practice, one needs to adopt a parametrized functional form for the control ansatz, leaving several free parameters to be determined numerically to minimize the relevant constraints. 
Recent experiments based on numerically optimized adiabatic ramps further highlight this complexity~\cite{Lancellotti2024, Marinelli2020}, underscoring the need for control strategies that extend beyond current constructions. 
While fully numerical approaches, such as optimal control methods~\cite{Weiss2021, Stefanatos2017}, provide one possible route, hybrid STA–numerical strategies offer the possibility of exploiting the analytical insight provided by STA techniques while reducing the complexity of the numerical optimization task.

In this paper, we bridge the gap by systematically evaluating the performance of various numerical optimization tools in optimizing the STA for the complex 2-ion separation problem.
We show that, within the commonly used normal-mode approximation of the Hamiltonian~\cite{James1998}, a wide range of numerical methods rapidly converge to similar solutions (free parameters).
However, once the anharmonic term is included, the complexity of the problem increases dramatically, making the choice of the optimization method crucial.
We demonstrate that although different numerical approaches do not converge to identical solutions, they nonetheless provide complementary insights that significantly reduce the effective search space.
Leveraging these numerical insights, our analytical-numerical approach identifies new sets of free parameters that yield orders-of-magnitude performance improvements.
In addition, these free parameters yield experimental control functions that are no more difficult to implement than those previously reported in the literature~\cite{Palmero2015}.

The remainder of this paper is structured as follows. 
Section~\ref{sec:control} formulates the control problem. 
Section~\ref{sec:harmonic} analyzes the optimization behavior within the harmonic approximation, and Sec.~\ref{sec:anharmonic} explores the algorithmic performance when anharmonic terms are included.
Finally, Sec.~\ref{sec:conclusion} summarizes the main results.

\section{The Control Problem}\label{sec:control}
The ion separation problem requires going from a single well configuration, with both ions trapped together, to a double well configuration, with each ion stored in a different trap. Typically, due to Paul trap symmetries canceling odd terms~\cite{Nizamani2011}, the external potential for a particle is defined as
\begin{equation}
    V_{ext}=\alpha(t) q^2+\beta(t) q^4,
\end{equation}
where in the initial single-well configuration the harmonic term $\alpha$ is positive and the quartic term $\beta$ is 0, and for the final double-well configuration $\alpha<0$ and $\beta>0$. For two interacting ions, the full Hamiltonian is then:
\begin{equation}\label{labHamiltonian}
    H = \frac{p_1^2}{2m}+\frac{p_2^2}{2m}+\alpha (t) (q_1^2+q_2^2)+\beta (t) (q_1^4+q_2^4)+\frac{C_c}{q_1-q_2},
\end{equation}
where $p_1, p_2$ are the momentum operators for both ions, $q_1, q_2$ their corresponding position operators, and $C_c=\frac{e^2}{4\pi\varepsilon_0}$, $\varepsilon_0$ being the vacuum permittivity. 
Following the procedure in~\cite{Palmero2015}, we rewrite the Hamiltonian~\eqref{labHamiltonian} after taking the dynamical normal mode~\cite{Lizuain2017} approximation around the equilibrium position $q_1^{(0)} = d(t)/2, q_2^{(0)} = -d(t)/2$, where $d(t)$ has to satisfy the quintic equation~\cite{Home2006}
\begin{equation}
    \beta(t)d^5(t)+2\alpha(t)d^3(t)-2C_c=0.
\end{equation}

After the normal mode transformation and a further unitary transformation (we refer to~\cite{Palmero2015} for full details), we reach the Hamiltonian
\begin{align}\label{HamNM}
    H_{NM} = H_+ + H_- &= \frac{{\sf p}_+^2}{2}+\frac{1}{2}\Omega_+^2\left({\sf q}_++\frac{\sqrt{m}\ddot{d}}{\sqrt{2}\Omega_+^2}\right)^2\nonumber\\
&+ \frac{{\sf p}_-^2}{2}+\frac{1}{2}\Omega_-^2{\sf q}_-^2,
\end{align}
where the $\dot{}$ notation indicates time derivative, ${\sf p}_{\pm}, {\sf q}_{\pm}$ are the conjugate position and momentum operators for both normal modes, and the $\Omega_{\pm}$ are their corresponding (time-dependent) frequencies given by
\begin{align}
    \Omega- &= \sqrt{\frac{1}{m}(2\alpha +3\beta d^2)},\nonumber\\
    \Omega_+ &= \sqrt{\frac{1}{m}\left(2\alpha +3\beta d^2+\frac{4C_c}{d^3}\right)}.
\end{align}

Since we will solve the dynamics within this normal mode approximation, we can conversely obtain the control parameters
and the related function $d(t)$
as a function of the normal mode frequencies with equations
\begin{align}\label{lab_parameters}
    d(t) &= \sqrt[3]{\frac{4C_c}{m\left(\Omega_+^2-\Omega_-^2\right)}},\nonumber\\
\alpha(t) &= \frac{1}{8}m\left(3\Omega_+^2-5\Omega_-^2\right),\nonumber\\
\beta(t) &= \frac{2C_c}{d^5(t)}-\frac{2\alpha(t)}{d^2(t)}.
\end{align}

The Hamiltonian~\eqref{HamNM} is made of two decoupled harmonic oscillators, so it has a well-known invariant of motion~\cite{Lewis1969}, provided we satisfy the auxiliary equations
\begin{align}
    \label{auxiliaryerm}
    \ddot{\rho}_\pm+\Omega_\pm^2\rho_\pm &= \frac{\Omega_{0\pm}^2}{\rho_\pm^3},
    \\
    \label{auxiliarynew}
    \ddot{x}_++\Omega_+^2x_+ &= -\sqrt{\frac{m}{2}}\ddot{d},
\end{align}
with $\Omega_{0\pm}=\Omega_\pm (0)$, and, due to symmetry, $x_-=0$. To guarantee the Hamiltonian leads us to the desired state, we impose the following boundary conditions (refer to~\cite{Palmero2015} and~\ref{appendix:invariant} for more details):
\begin{align}
    \rho_\pm(0) &= 1,\;\; \rho_\pm(t_f)=\gamma_\pm, 
    \label{rho1}
    \\
    \dot{\rho}_\pm(t_b) &= \ddot{\rho}_\pm(t_b)=0,
    \label{rho2}
    \\
    \label{xbc}
    x_+(t_b) &= \dot{x}_+(t_b)=\ddot{x}_+(t_b)=0,
    \\
    \label{rho3}
    \dddot{\rho}_\pm(t_b) &= \ddddot{\rho}_\pm(t_b)=0,
\end{align}
where $\gamma_\pm = \sqrt{\frac{\Omega_\pm(0)}{\Omega_\pm(t_f)}}$, and $t_b$ stands for both boundary times, $0; t_f$.
It is impossible to analytically find an ansatz that satisfies all these boundary conditions simultaneously.
Instead, we begin with an ansatz that satisfies as many as possible, and leave free parameters that are to be numerically fixed to minimize the excitation energy. For comparison purposes, we use a polynomial ansatz of up to degree 12; $\rho_+ = \sum_{i=0}^{12} a_is^i$ ($s=t/t_f$), leaving up to three free parameters, as in~\cite{Palmero2015}.
Full details of this ansatz can be seen in the~\ref{appendix:invariant}. 

To fix the free parameters, we will define two different cost functions.
First, staying within the harmonic approximation of the normal mode approximation, the cost function will simply be the excitation energy of Hamiltonian~\eqref{HamNM} at the final time
\begin{align}\label{cost_function}
    F &= E_{n+} + E_{n-} = \frac{(2n+1)\hbar}{4\Omega_{0-}}\left(\dot{\rho}_-^2+\Omega_-^2\rho_-^2+\frac{\Omega_{0-}^2}{\rho_-^2}\right)\nonumber\\
    &+ \frac{(2n+1)\hbar}{4\Omega_{0+}}\left(\dot{\rho}_+^2+\Omega_+^2\rho_+^2+\frac{\Omega_{0+}^2}{\rho_+^2}\right) + \frac{1}{2}\dot{x}_+^2+\frac{1}{2}\Omega_+^2\left(x_+-\frac{\sqrt{m}\ddot{d}}{\sqrt{2}\Omega_+^2}\right)^2.
\end{align}

The second approach will additionally consider the next term (cubic) previously neglected in the normal mode approximation, and try to minimize it simultaneously with the cost function.
Following~\cite{Morigi2001}, in terms of our normal mode notation, the higher-order anharmonicities lead to the terms
\begin{equation}
    \delta V^{(j)}=(-1)^{j+1}\frac{C_c}{d^{j+1}}\left(\sqrt{\frac{2}{m}}{\sf q}_+\right)^j,
\end{equation}
which leads to an additional excitation energy,
\begin{equation}
    \delta E_{n+}^{(j)}=\langle \psi_{n+}(t_f)|\delta V^{(j)}|\psi_{n+}(t_f)\rangle,
\end{equation}
where the $\psi_{n+}$ is the eigenstate of the "+"-mode part of the normal-mode Hamiltonian~\eqref{HamNM}.
Since we will only keep cubic terms, our new cost function will be 
\begin{equation}\label{cost_function_qubic}
    F_{qub} = F + \epsilon\delta E_{0+}^{(3)},
\end{equation}
where $\epsilon=10^{-8}$ is a weight we manually introduce to the anharmonic term – fixed after a trial and error process\footnote{
Specifically, we tested a wide range of values for the parameter $\epsilon$ at several final times and selected the value that minimized the final excitation energy.
}. 
This new term is far from trivial, numerically speaking.
It involves evaluating an integral for each set of values we try out for the free parameters in our ansatz.
As we will show in the following sections, opposite to the purely harmonic cost function, the different numerical methods can lead to very different solutions, with varying degrees of minimization. 

\section{Results I: Harmonic Approximation}\label{sec:harmonic}
\begin{figure}[htp!]
    \centering
    \includegraphics[width=\linewidth]{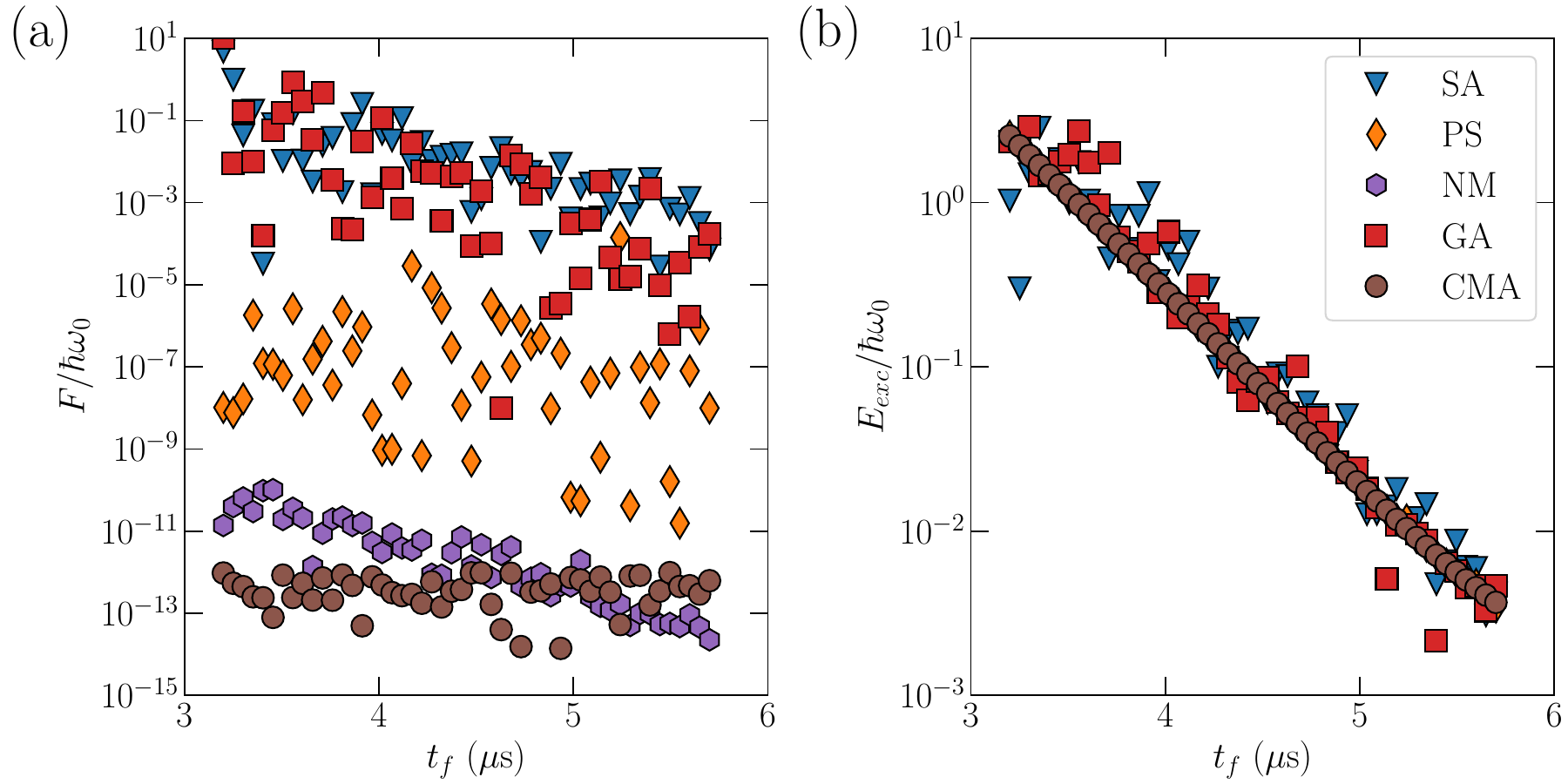}
    \caption{(a) The optimal value of the cost function $F/\hbar \omega_0$ versus final time and (b) the actual final excitation energy $E_{exc}/\hbar \omega_0$ versus final time, for all numerical methods considered.
    The simulations are done for two $^9\text{Be}$ ions, $\alpha(0) = \frac{1}{2}m\omega_0^2$, where $m=9$ AMU, $\omega_0 = 2$ MHz, $\alpha({t_f}) = - \frac{1}{2} \alpha(0)$, and $d(t_f) = 10 d(0)$.}
    \label{fig:Harmonic_E}
\end{figure}

We apply various numerical optimization methods to the harmonic approximation of the above 2-ion separation problem and work with constants that map directly to experiments.
The numerical methods of choice are simulated annealing (SA)~\cite{Ingber1996}, particle swarming (PS)~\cite{Kennedy1995}, Nelder-Mead (NM)~\cite{Lagarias1998}, genetic algorithm (GA)~\cite{Goldberg1988}, and covariant matrix adaptation evolution strategy (CMA)~\cite{Hansen2001}. 
The simulations are performed for two $^9\text{Be}$ ions, $\alpha(0) = \frac{1}{2}m\omega_0^2$, where $m=9$ AMU, $\omega_0 = 2$ MHz, $\alpha({t_f}) = - \frac{1}{2} \alpha(0)$, and $d(t_f) = 10 d(0)$.
For this first set of optimizations, we find by trial and error that only two free parameters suffice to converge towards the optimal solution, so we fix the 12th parameter in our polynomial to $a_{12}=0$, and only optimize with respect to free parameters $a_{10}$ and $a_{11}$.

In Fig.~\ref{fig:Harmonic_E}, we plot (a) the optimal value of the cost function $F$ (Eq.~\eqref{cost_function}) versus final time.
We see a large variation in the convergence accuracy of the different numerical methods, with CMA and NM consistently outperforming the others for most $t_f$ considered.
However, when looking at the final excitation energy $E_{exc}$ of the real problem (evolving with the full Hamiltonian~\eqref{labHamiltonian}) in Fig.~\ref{fig:Harmonic_E}(b), we see that the optimal control parameters found by most numerical methods lead to similar actual final excitation energy, except for SA and GA, which tend to fluctuate from run to run.
Therefore, the difference in $F$ for different methods is largely due to the fine-tuning in the precision of the free parameters $a_{10}$ and $a_{11}$.

\begin{figure}[htp!]
    \centering
    \includegraphics[width=\linewidth]{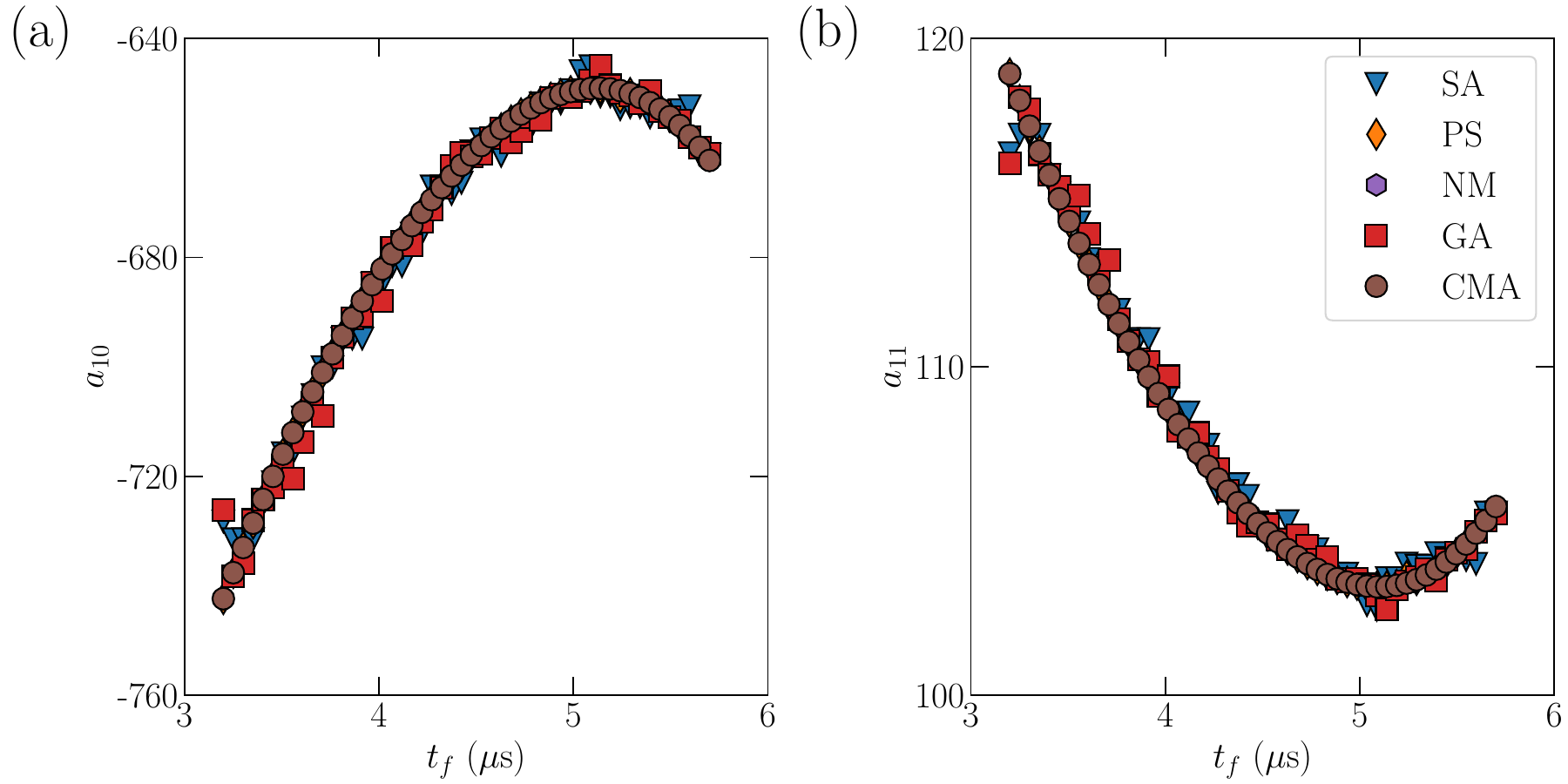}
    \caption{The optimal optimization parameter (a) $a_{10}$ and (b) $a_{11}$ versus final time found by different numerical methods for the harmonic case. Same parameters as in Fig.~\ref{fig:Harmonic_E}.}
    \label{fig:Harmonic_a}
\end{figure}

In Fig.~\ref{fig:Harmonic_a}, we show that the optimal free parameters $a_{10}$ and $a_{11}$ found by different numerical methods are indeed very similar.
This indicates that the optimization problem in the harmonic approximation is relatively straightforward, and most numerical methods can find similar optimal free parameters.
One area of concern, however, is that the values of $a_{10}$ and $a_{11}$ found here are unbounded, which may lead to difficulties in convergence as the optimization complexity and parameter space increase.

\section{Results II: Anharmonic Approximation}\label{sec:anharmonic}
\begin{figure}[htp!]
    \centering
    \includegraphics[width=\linewidth]{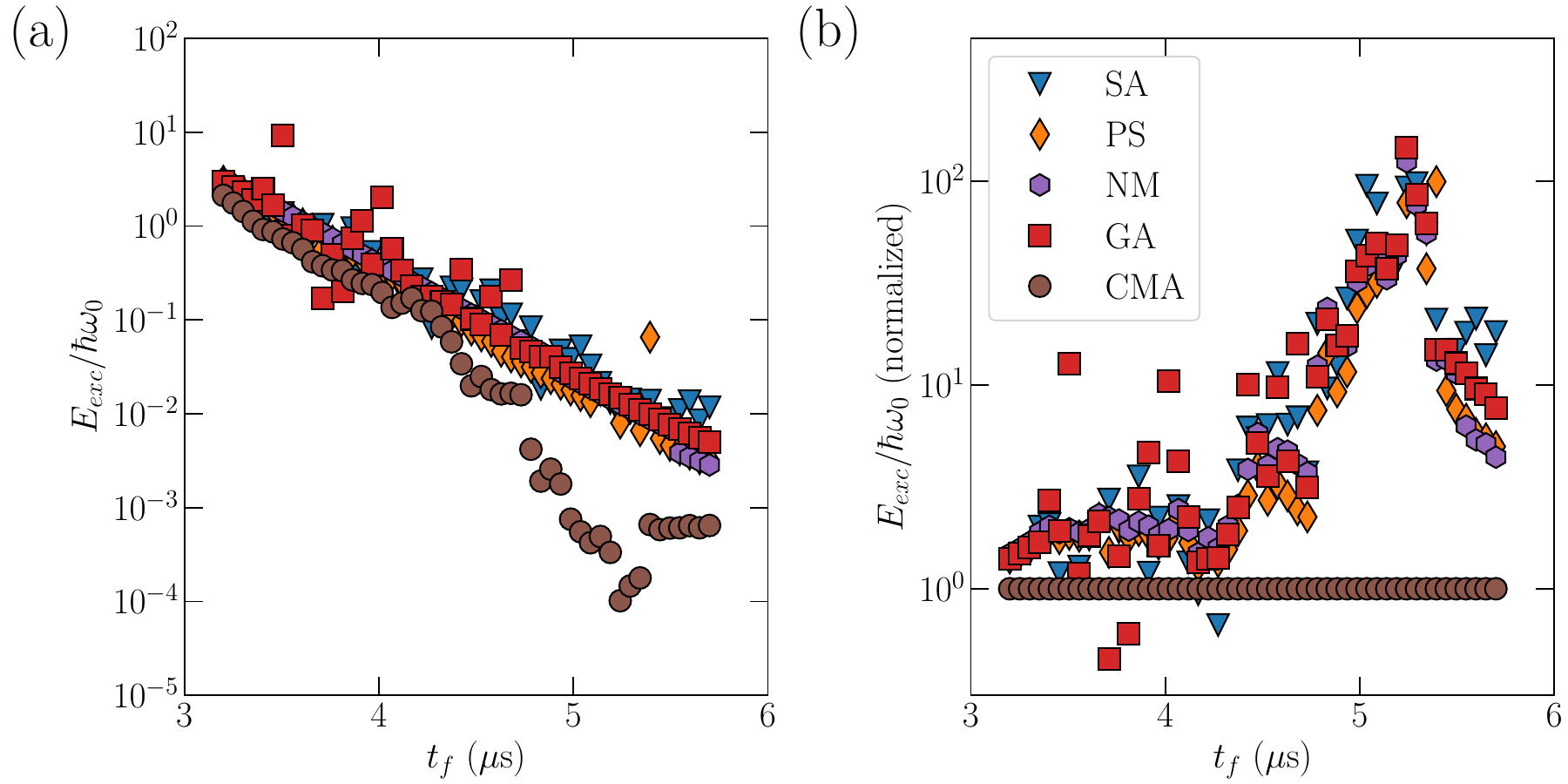}
    \caption{(a) The actual final excitation energy $E_{exc}$ versus final time, and (b) the normalized final excitation energy $E_{exc}/E_{exc}^{\text{CMA}}$ versus final time, for all numerical methods considered in the cubic case. Same parameters as in Fig.~\ref{fig:Harmonic_E}.}
    \label{fig:Qubic_E_normalized}
\end{figure}

When considering the additional anharmonic term, as described in Eq.\eqref{cost_function_qubic}, the optimization problem becomes significantly more difficult.
In Fig.~\ref{fig:Qubic_E_normalized}(a), we plot the actual final excitation energy $E_{exc}$ obtained from the protocols given by the different numerical methods.
Unlike in the harmonic approximation, we see a large variation in $E_{exc}$ obtained by different numerical methods, especially at longer $t_f$.
With the anharmonic term, CMA consistently outperforms other numerical methods, sometimes by orders of magnitude.
To better visualize the performance of different numerical methods, in Fig.~\ref{fig:Qubic_E_normalized}(b), we plot the normalized final excitation energy $E_{exc}/E_{exc}^{\text{CMA}}$, where $E_{exc}^{\text{CMA}}$ is the final excitation energy obtained by the CMA method at the same $t_f$.

\begin{figure}[htp!]
    \centering
    \includegraphics[width=\linewidth]{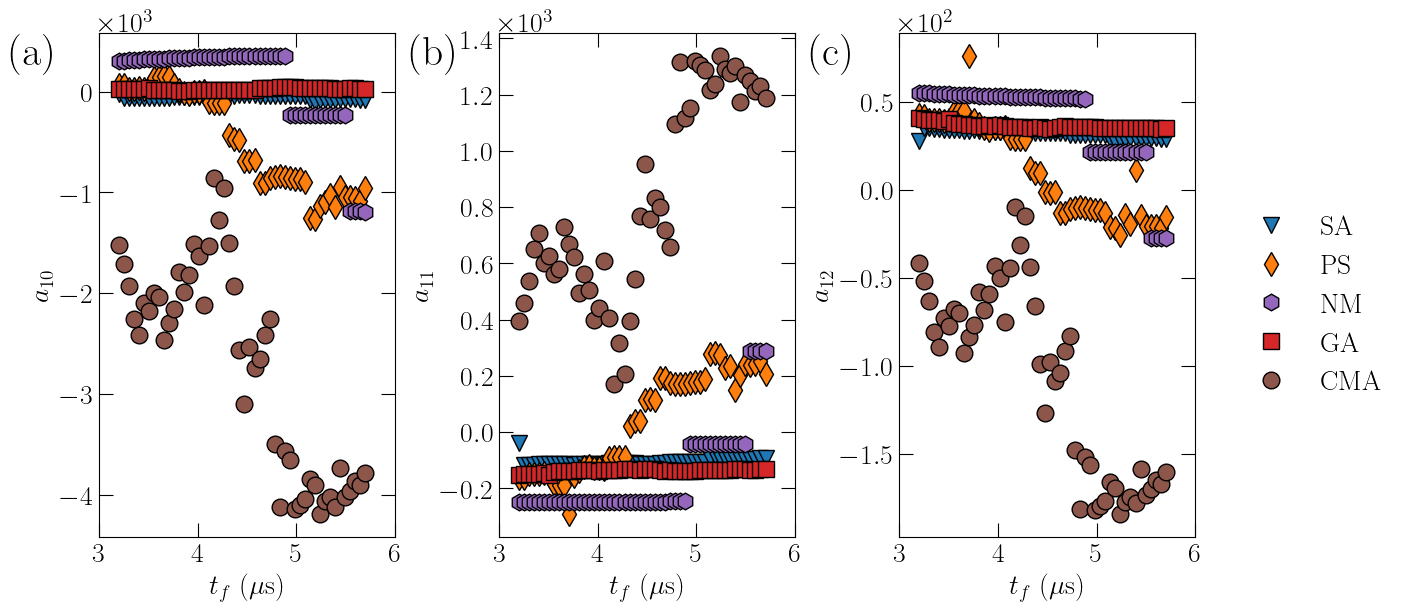}
    \caption{The optimal optimization parameter (a) $a_{10}$, (b) $a_{11}$, and (c) $a_{12}$ versus final time found by different numerical methods in the cubic case. Same parameters as in Fig.~\ref{fig:Harmonic_E}.}
    \label{fig:Qubic_a}
\end{figure}

One common reason for different numerical methods yielding alternative results is that they are sometimes trapped in other local minima in the complex parameter space.
To show this, we plot the optimal free parameters (a) $a_{10}$, (b) $a_{11}$, and (c) $a_{12}$ found by different numerical methods in Fig.~\ref{fig:Qubic_a}.
We observe that CMA is the only numerical method that is capable of exploring a wide range of parameter space, while other methods tend to cluster within a much narrower parameter space.
As mentioned earlier, the free parameters have no physical meaning, so there is no reason to believe that the optimal solution should be bounded within a certain range.
The CMA method converged to much better $E_{exc}$ values because it explored a much wider range of optimization parameters.

We can leverage this insight to manually expand the parameter space and search for free parameters that give even lower final excitation energy.
However, expanding the parameter search space via a brute-force grid search is not feasible because of the dimensionality, extremely high-precision requirement, and the unbounded nature of the free parameters.
If we can deduce some patterns in the local minima found by different numerical methods, we can extrapolate this pattern to find new optimization parameters that may yield even better performance.

\begin{figure}[htp!]
    \centering
    \includegraphics[width=0.7\linewidth]{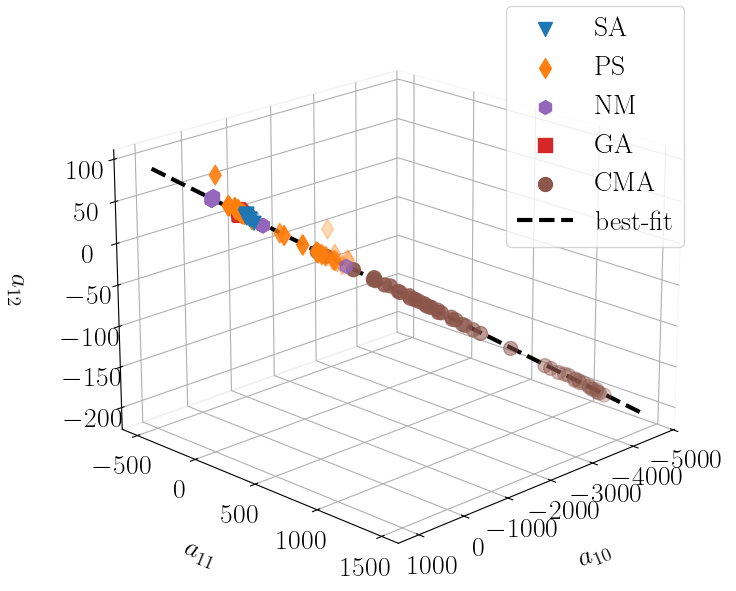}
    \caption{$3$D plot of all optimized solutions (for all methods and all final times) shown in Fig.~\ref{fig:Qubic_a}, along with the linear fit curve that best represents them.}
    \label{fig:all_solutions_3D}
\end{figure}

In Fig.~\ref{fig:all_solutions_3D}, we collect the free parameters $a_{10}$, $a_{11}$, and $a_{12}$ (for all $t_f$) found by different numerical methods in a 3D plot.
Interestingly, we observe that a majority of these points approximately lie on a straight line in this $3$D space.
If another value of the weighing parameter $\epsilon$ in Eq. \eqref{cost_function_qubic} is chosen, the results obtained there would still lie on the same straight line we observe in Fig.~\ref{fig:all_solutions_3D}, but often with a higher final excitation energy.

Given this line of local minima, the optimization problem is reduced to a $1$D search along this line.
We compute a center point $\bar{a}$ by averaging all the points shown in Fig.~\ref{fig:all_solutions_3D} and define a line direction vector $\vec{v}$ via a linear fit.
We then sample points $a^\prime = \bar{a} + \nu \vec{v}$ along this line direction vector by varying $\nu$ and use the $a^\prime$ as the starting seeds to perform NM optimization.

\begin{figure}[htp!]
    \centering
    \includegraphics[width=0.7\linewidth]{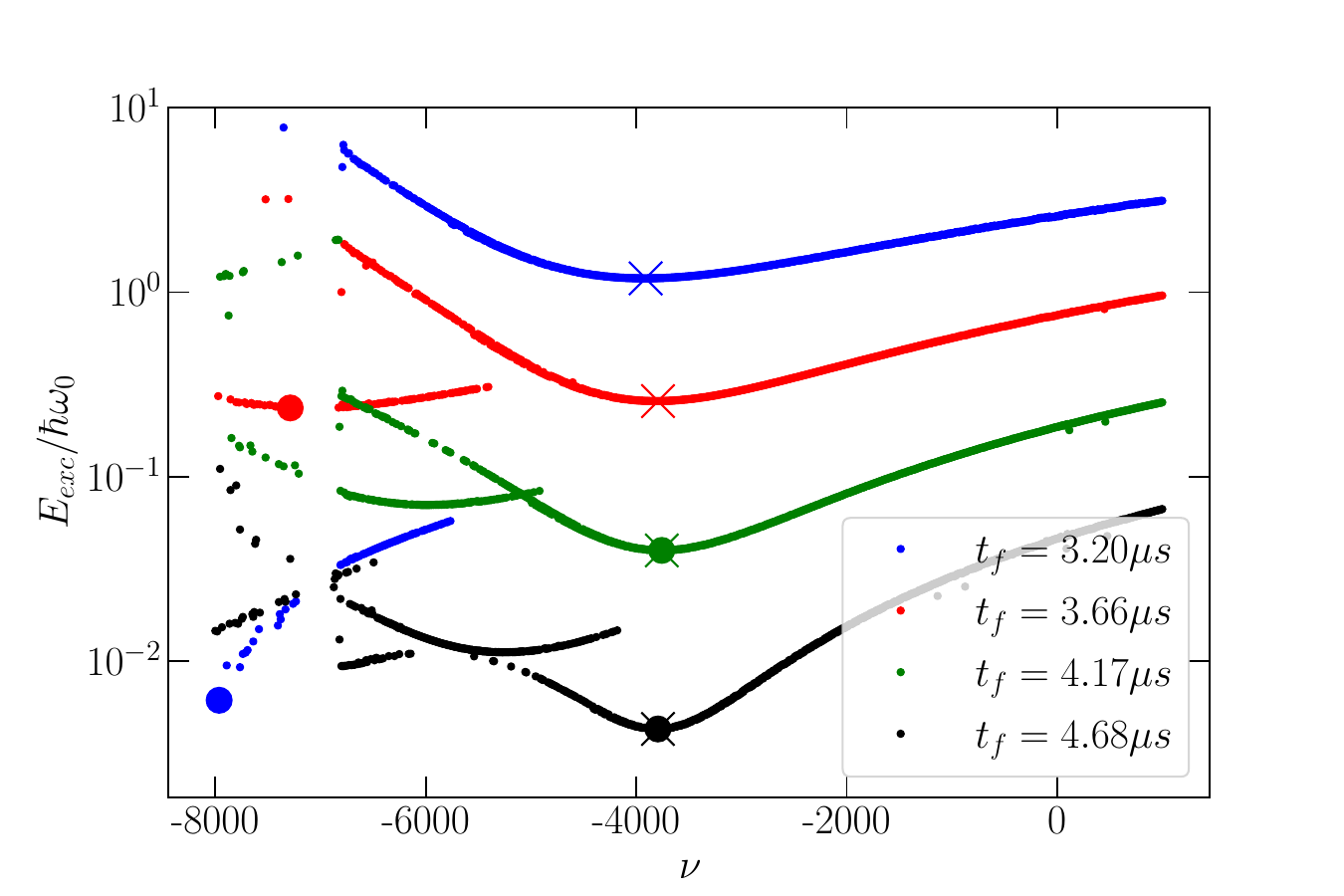}
    \caption{Actual excitation energy at different final times after optimizing with the NM method vs different starting values of $\nu$ along the line of optimal values chosen as a starting seed. Same parameters as in Fig.~\ref{fig:Harmonic_E}.}
    \label{fig:Qubic_line}
\end{figure}

In Fig.~\ref{fig:Qubic_line}, we plot the final excitation energy $E_{exc}$ obtained by using points along the line as the initial condition and for four final times, $t_f = 3.20, 3.66, 4.17$ and $4.68\mu$s.
%, $t_f = 3.66$~$\mu$s, $t_f = 4.17$ $\mu$s, and $t_f = 4.68$ $\mu$s.
For all times, we find two $\nu$ regions: at lower $|\nu|$, the final excitation energy $E_{exc}$ changes smoothly with $\nu$, while at higher $|\nu|$, $E_{exc}$ fluctuates discontinuously with $\nu$ (at some $\nu$ values the NM optimization fails to converge).
We indicate the best (minimum) $E_{exc}^{\text{best}}$ with a filled circle and the local (minimum) $E_{exc}^{\text{local}}$ in the smooth region with an ``X'' symbol.
At longer $t_f$, $E_{exc}^{\text{best}}$ and $E_{exc}^{\text{local}}$ tend to be at the same $\nu$ value.
At shorter $t_f$, we find that $E_{exc}^{\text{best}}$ is found outside of the smooth region.
The discontinuity of $E_{exc}$ beyond the smooth region suggests that the optimization landscape is highly complicated, and small changes in the free parameters can lead to large changes in the final excitation energy. 
The problems encountered here are twofold: the optimal solution valley is very narrow and rugged, requiring extreme precision and sensitivity, and there are unphysical regions in the vicinity of the smallest $\nu$ region (negative values for $d$) where the mathematical problem returns complex values, returning errors in the optimization process.
Therefore, it is not surprising that heuristic and metaheuristic methods struggle to find the optimal parameters in this complex landscape.

\begin{figure}[htp!]
    \centering
    \includegraphics[width=\linewidth]{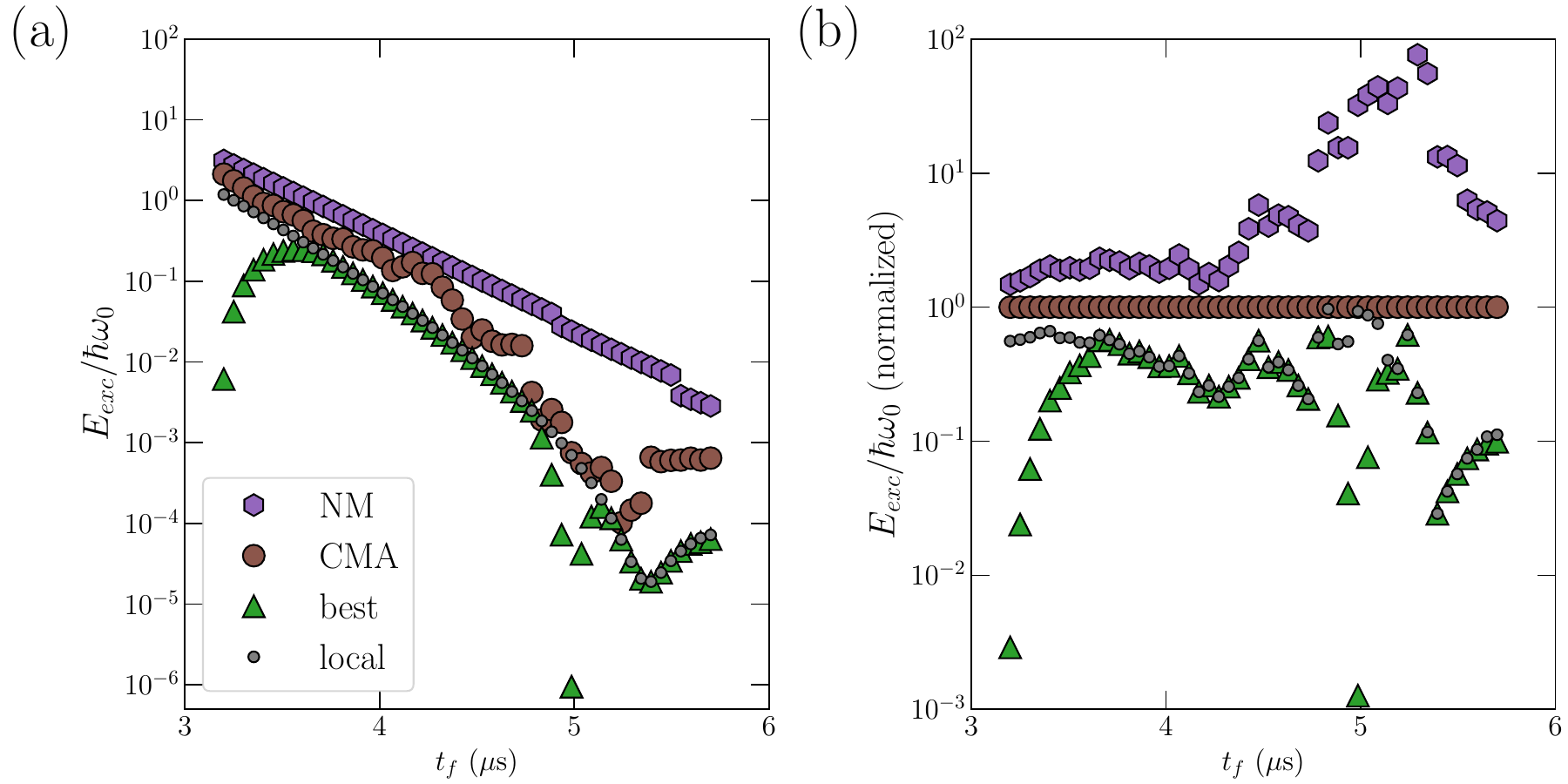}
    \caption{(a) Actual excitation energy and (b) the same normalized against the results for the CMA method $E_{exc}/E_{exc}^{\text{CMA}}$. The figures show the absolute best, and locally best solutions found when optimizing along the extended line of optimal solutions (as done in Fig.~\ref{fig:Qubic_line}), and compare them to the benchmark CMA and NM solutions from the simple approach.}
    \label{fig:Qubic_E_best_normalized}
\end{figure}

In Fig.~\ref{fig:Qubic_E_best_normalized}, we plot $E_{exc}^{\text{best}}$ and $E_{exc}^{\text{local}}$ obtained by this line search method as a function of $t_f$, and compare them to the $E_{exc}$ obtained by the CMA method.
Both $E_{exc}^{\text{best}}$ and $E_{exc}^{\text{local}}$ outperform the CMA results for all $t_f$ considered.
In particular, $E_{exc}^{\text{best}}$ outperforms CMA by almost three orders of magnitude at the shortest $t_f$, which is the regime of most interest for experimental realizations.

\begin{figure}[htp!]
    \centering
    \includegraphics[width=\linewidth]{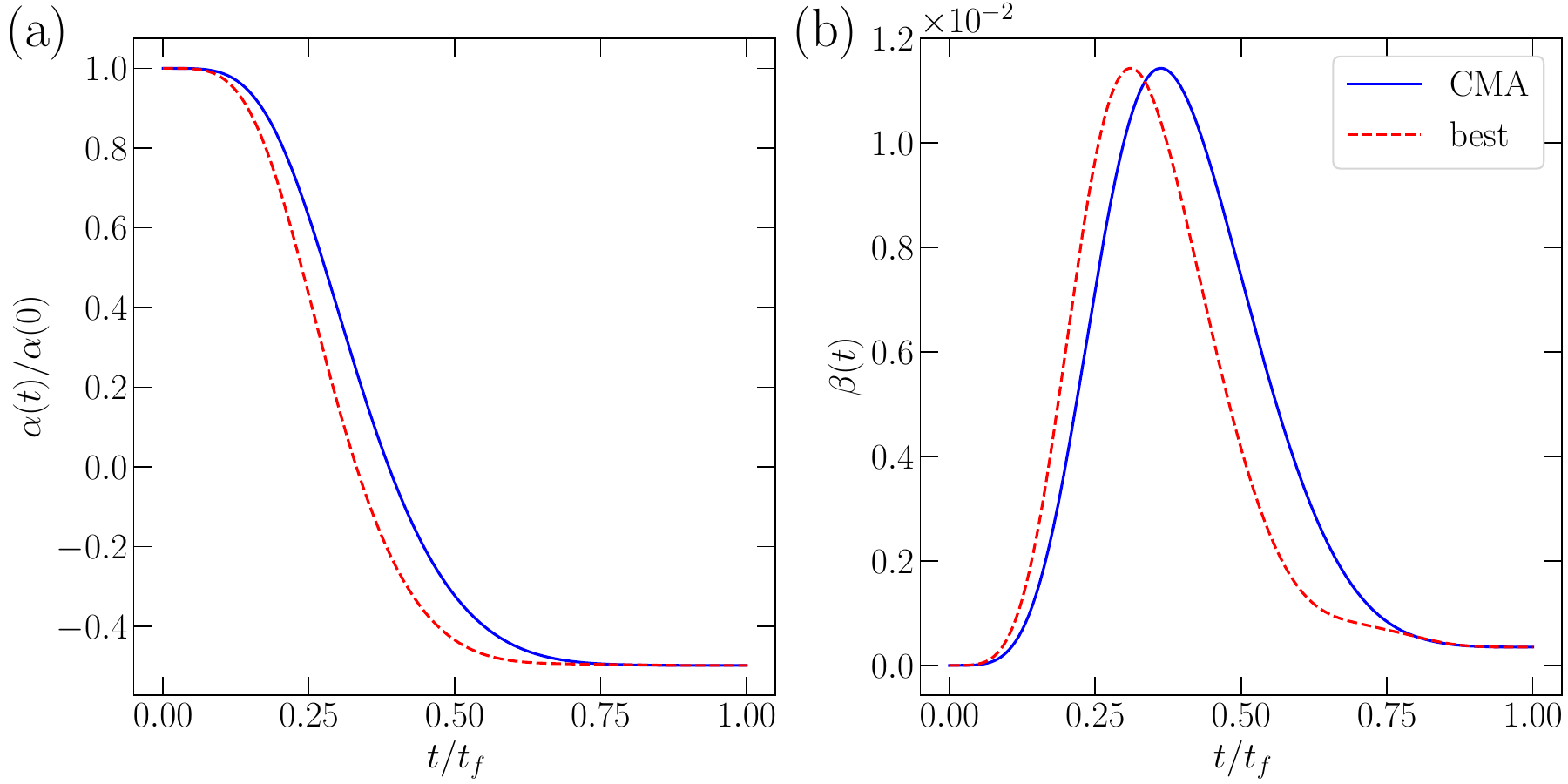}
    \caption{(a) The lab parameter $\alpha$ and (b) $\beta$ vs time for the best solution for the CMA and the absolutely best solution along the optimal line at $t_f = 3.2$ $\mu$s. Same parameters as in Fig.~\ref{fig:Harmonic_E}.}
    \label{fig:Qubic_E_control}
\end{figure}

Lastly, we show that the optimal control function found by this line search methods are not more difficult to implement experimentally compared to the previous ones found in the literature~\cite{Palmero2015}.
We calculate the control functions $\alpha(t)$ and $\beta(t)$ using the optimal parameters $a^{\text{best}}$ found at $t_f = 3.20$~$\mu$s and compare them to $a^{\text{CMA}}$ found by the CMA method at the same $t_f$.
The control function for both $\alpha(t)$ and $\beta(t)$ are shown in Fig.~\ref{fig:Qubic_E_control}(a) and (b), respectively.
We observe that the control functions obtained by both methods are of similar magnitude and smoothness.
At the same time, the control functions are distinguishable.
Therefore, the improved control function we found can be implemented in experiments without significant modifications to the existing setup. 
Particularly, all methods, with all levels of optimization, lead to the same maximum value of the $\beta$ parameter, a known experimental roadblock towards achieving faster driving protocols~\cite{Palmero2015}.
This suggests that the maximum value for $\beta$ is a hard physical requirement, and optimizing the choice of the free parameters can lead to orders-of-magnitude improvements without increasing technical requirements.

\begin{figure}[htp!]
    \centering
    \includegraphics[width=\linewidth]{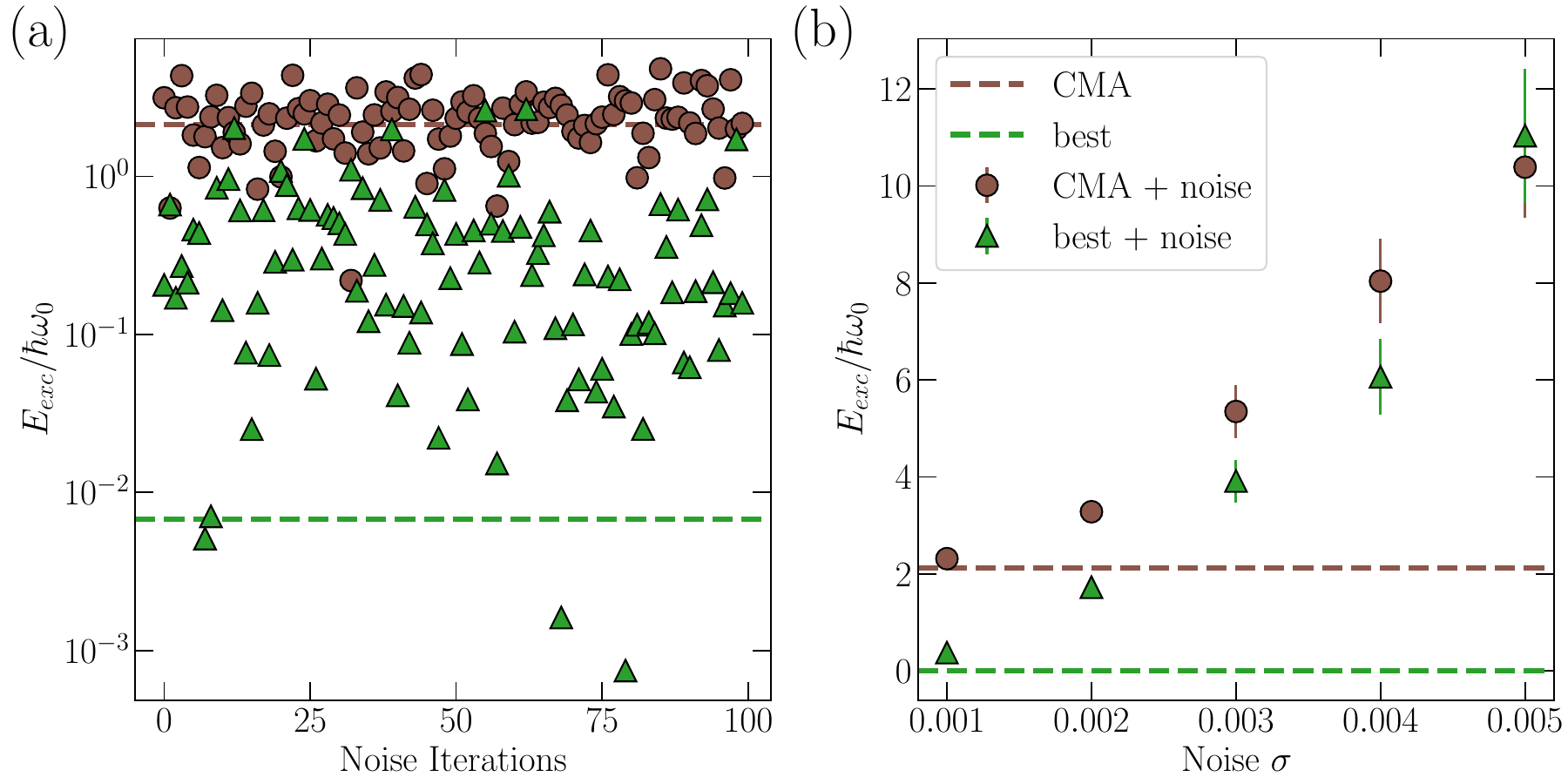}
    \caption{(a) $E_{exc}$ obtained from $100$ noisy implementations of the CMA control functions and the best control function.
    The noisy control functions are obtained by applying a Gaussian factor, $\alpha(t)^\prime = \alpha(t) \times N(1, \sigma)$, $\beta(t)^\prime = \beta(t) \times N(1, \sigma)$, where $\sigma = 0.001$.
    (b) $E_{exc}$ as a function of $\sigma$ for both the CMA control function and the best control function.
    The dashed lines are the reference $E_{exc}$ produced by the exact CMA and best control functions.}
    \label{fig:noise}
\end{figure}

In Fig.~\ref{fig:noise}, we show that the control function obtained from $a^{\text{best}}$ performs better than that from $a^{\text{CMA}}$ even in the presence of small experimental imperfection.
In Fig.~\ref{fig:noise}(a), we add a Gaussian noise ($\sigma = 0.001$) to the control functions $\alpha(t)^\prime = \alpha(t) \times N(1, \sigma)$, $\beta(t)^\prime = \beta(t) \times N(1, \sigma)$, and compute $E_{exc}$ from $100$ different iterations of this noisy control function.
It is evident that, even with small imperfections in experiments, the control functions derived from $a^{\text{best}}$ outperforms that from $a^{\text{CMA}}$.
In Fig.~\ref{fig:noise}(b), we increase $\sigma$ and find that the control function obtained from $a^{\text{best}}$ continues to do better than $a^{\text{CMA}}$ at substantial noise $\sigma \leq 0.004$. At this point, the excitations obtained from either protocol are much higher than in the ideal case. 
This suggests that, although the best solution might be in a slightly less stable region, so long as we can achieve a sufficiently good control in the lab for any interesting application, this ramp will still outperform those achieved from simpler approaches.

\section{Conclusions}\label{sec:conclusion}

In this work, we apply various numerical optimization methods to the STA protocol for fast two-ion separation.
While most numerical methods can find similar optimal solutions under the harmonic approximation, the inclusion of anharmonic terms significantly increases the complexity of the optimization problem.
Among the heuristic and metaheuristic methods we have explored, CMA is often the most suitable optimization method for this class of problems.
However, by analyzing the set of solutions found by different numerical methods, we have further identified better free parameters that yield orders-of-magnitude lower final excitation energy improvement.
These new solutions also lead to experimental control functions that are not more difficult to implement compared to previous works.

In recent years, there has been a growing interest in applying machine learning techniques to quantum control~\cite{Ding2023, Perret2024, Xu2024} or optimization~\cite{Levine2014, Andrychowicz2016} problems.
While it is a promising direction to explore the application of machine learning techniques to the STA protocol, we note that it is not straightforward to apply machine learning techniques directly.
Firstly, solutions to the STA problem are found by solving a set of boundary value problems, which are currently not gradient preserving in most machine-learning frameworks.
It is possible to reframe the boundary value problem as an initial value problem, but the computational cost of computing the gradient of the final excitation energy with respect to the free parameters is extremely high.
Secondly, the optimization landscape of the STA problem contains many razor-thin local minima and large unphysical regions.
Lastly, the free parameters in the STA ansatz are unbounded and highly sensitive to fine-tuning at high precision.
Although a different parameterization of the STA ansatz might ease the optimization, we choose to work directly with experimentally meaningful parameters to ensure straightforward translation to the experimental platform.

While this work has focused on the specific problem of two-ion separation, the insights and techniques developed here can be readily extended.
In many optimization problems, “valleys” of solutions can emerge in high-dimensional and unbounded parameter spaces. Most numerical optimization methods can find these “valleys” but may get stuck in one of the local minima within them. This could be due to the exceptionally complex and sharp landscape within the “valleys”. The line observed in this manuscript is one example of such a “valley”, and it contains a large number of local minima. However, by using different numerical methods, some with better ability to explore or converge, one can identify different local minima, which gives information on the geometric properties of the “valleys”. The knowledge of such geometric properties can greatly improve our ability to find the global minimum.

The STA method of choice, the invariant-based inverse engineering, allows ample flexibility in the layout of the problem, including additional optimization requirements (as we did here by adding the cubic term). A particularly interesting set of problems where the hybrid approach practiced here could have a real impact is considering (and trying to mitigate) the effect of noise and errors~\cite{Ruschhaupt2012, Lu2013, Lu2014}. 
Another relevant set of problems is when we have physical limitations in the lab parameters. E.g., in the problem studied here, the maximum value of $\beta$ along the driving can be a concern, but the analysis shows that all methods converge towards the same requirement.
In other situations, exploring optimal regions could lead to different parameter requirements at the cost of small additional excitations.
Our work establishes a framework for applying and interpreting numerical optimization techniques in quantum control problems characterized by complex parameter landscapes.

\section*{Acknowledgements}
We acknowledge the helpful discussion with William Sharpless and Boning Li.
We are grateful for funding from the Spanish Ministry of Science, Innovation, and Universities through projects PID-2021-126277NB-I00, PID2024-156808NB-I00,
the Basque Government through project IT1470-22 and KUBIBIT project (KK-2025/00079) of Elkartek program,
the UPV/EHU through project EHU-N25/38,
the EU Flagship on Quantum Technologies through OpenSuperQ+100 (Grant No. 101113946),
and the National Science Foundation under Grant No. PHY2317134 (Center for Ultracold Atoms).
B. X. is supported by the Agency for Science, Technology and Research (A*STAR) International Fellowship. 
M. P. is supported by the Spanish Ministry of Science, Innovation and Universities through the Jos\'e Castillejo mobility grant (CAS24/00344).

\bibliographystyle{elsarticle-num}
\bibliography{bibliography}

\appendix
\section{More details on the invariant problem}\label{appendix:invariant}
A Hamiltonian with a harmonic form has a well-known invariant of motion introduced by Lewis and Riesenfeld~\cite{Lewis1969}. Adapting it to the two independent harmonic oscillators of the same form of each of the $H_\pm$ in Eq.~\eqref{HamNM}, the invariant for our case takes the form
\begin{align}
I_\pm &= \frac{1}{2}[\rho_\pm ({\sf p}_\pm-\dot{x}_\pm)-\dot{\rho}_\pm({\sf q}_\pm-x_\pm)]^2 + \frac{1}{2}\Omega_{0\pm}^2\left(\frac{{\sf q}_\pm-x_\pm}{\rho_\pm}\right)^2,
\end{align}
where $\rho_\pm$ and $x_+$ are auxiliary functions that have to satisfy each a pair of auxiliary equations~\eqref{auxiliaryerm} and~\eqref{auxiliarynew}. 

The strategy followed in the invariant-based inverse engineering method is to impose commutativity between the Hamiltonian and the invariant at both boundary times. By definition, when these two commute, they share the same eigenstate, and since the invariant always follows the same one along the whole evolution, commutation at both boundary times guarantees that the Hamiltonian is driven back to the same state it was initialized in at the end of the evolution, although we have no control over what happens at all intermediate times. These commutations are guaranteed by the set of boundary conditions presented in Eqs.~\eqref{rho1},~\eqref{rho2} and~\eqref{xbc}.

Additionally, for smoothness in the unitary transformations that led to the Hamiltonian in Eq.~\eqref{HamNM}, we impose $\dot{d} (t_b) = \ddot{d}(t_b) = 0$, which amounts to imposing the additional boundary conditions in Eq.~\eqref{rho3}.

The strategy is now to begin with ans\"atze for the $\rho_{\pm}$ that satisfy all the corresponding boundary conditions, leaving some free parameters for the other boundary conditions. These ans\"atze are then introduced in Eq.~\eqref{auxiliaryerm}, where solving the equations, the $\Omega_\pm$ are obtained, from which we get the control parameters in Eq.~\eqref{lab_parameters}. These are then used back in the remaining auxiliary Eq.~\eqref{auxiliarynew}. We solve this differential equation and fix the free parameters to satisfy all the remaining boundary conditions. Unfortunately, the Ermakov equation~\eqref{auxiliaryerm} is particularly complicated due to the $\sim\frac{1}{\rho^3}$ term, which leads to divergences. In practice, it is not feasible to solve this differential equation analytically when some free parameters are left, and therefore, the only option left is to numerically solve and fix the free parameters. In~\cite{Palmero2015}, this procedure was followed using the polynomial ans\"atze we introduced in the main text. After fixing some parameters to meet the boundary conditions described in~\eqref{rho1},~\eqref{rho2}, and~\eqref{rho3}, and leaving up to three free parameters, the resulting ansatz is
\begin{align}
\label{ansatz}
    \rho_- &= 126(\gamma_--1)s^5 - 420(\gamma_--1)s^6 + 540(\gamma_--1)s^7\nonumber\\
    &- 315(\gamma_--1)s^8 + 70(\gamma_--1)s^9 + 1,\nonumber\\
    \rho_+ &= 1-(126-126\gamma_++a_{10}+5a_{11}+15a_{12})s^5\nonumber\\
    &+(420-420\gamma_++5a_{10}+24a_{11}+70a_{12})s^6\nonumber\\
    &- (540-540\gamma_++10a_{10}+45a_{11}+126a_{12})s^7\nonumber\\
    &+ (315-315\gamma_++10a_{10}+40a_{11}+105a_{12})s^8\nonumber\\
    &- (70-70\gamma_++5a_{10}+15a_{11}+35a_{12})s^9\nonumber\\
    &+ a_{10}s^{10}+a_{11}s^{11}+a_{12}s^{12}.
\end{align}
Other functional forms could be considered (Fourier cosines, Chebysev Polynomials...), and are often more efficient because they require fewer terms to meet all boundary conditions. However, in this paper, we stick to the simple polynomial for a fair comparison with~\cite{Palmero2015}.

\section{Details on the numerical optimization}\label{appendix:optimization}
In the main text, we described how we tried minimizing both the cost function for the harmonic case~\eqref{cost_function} and for the cubic case~\eqref{cost_function_qubic} using five different numerical methods: NM, GA, SA, PS, and CMA. All these are well-known heuristic and metaheuristic methods, so we implemented the minimization of pre-existing functions. NM, GA, SA, and PS have built in MATLAB functions, while for the CMA, we downloaded a MATLAB-ready code from~\cite{YarpizWEB}. For all methods, we followed similar strategies; using (0, 0, 0) or a homogeneous spread around this point as a starting seed for the shortest final time, and
letting the optimizer find the optimal solution in a single shot.
Subsequently using the previous optimal result for the following final times. Here are some additional details on how we specifically implemented each of the methods:
\begin{itemize}
    \item \textbf{Nealder-Mead} 
    We used the NM algorithm implemented in MATLAB's \emph{fminsearch} function, and set a relative tolerance of $10^{-7}$, and a maximum number of iterations of 5000, more than enough for all our tests. We started the search with a seed value of $(0,0,0)$ for the shortest final time, and the previous optimal solution as the subsequent seeds. 
    \item \textbf{Genetic Algorithms} 
    We set an initial population of 50 individuals, uniformly spread between $[-60, 60]$ for the first point in the final time, and for subsequent final times with a narrower spread of $\pm5$ around the previous optimal solution. We ranked the population purely based on the output value of the cost function. Parent selection was done in a tournament format, randomly selecting 5 members and choosing the fittest one. For the crossover, we chose the new value at some point in the straight line that links both parents, weighting it based on each parent's performance. Finally, for the mutation, we left the default settings. 
    \item \textbf{Particle Swarm Algorithm} 
    We implement it in MATLAB's \emph{particleswarm} function. Like for the GA, we set an initial population of 50, spread between $[-100, 100]$ for the initial point, and subsequently with a $\pm30$ spread around the previous optimal point. We set an inertia range $W\in [0.01, 1.40]$ –a parameter that indicates how similar the velocity parameter of a given subject is across generations– and self- and social-adjustment parameters are both set to 1.49 –how much we weight each particle's best position and the neighbors' best position to update the position parameter.
    \item \textbf{Simulated Annealing}
    We implement it using MATLAB's \emph{simulannealbnd} function. We choose an initial temperature $T_0 = 200$, and a reannealing interval of 10. After the first solution is found, the step size is lowered by square rooting it to incentivize sweeping the valley with a better resolution, and similarly, we change the annealing function from a traditional exponential to a hyperbolic one.
    \item \textbf{Covariance Matrix Adaptation Evolution Strategy}
    We implemented a function available through File Exchange~\cite{YarpizWEB}. We slightly modified this code, limiting the maximum step size to $10^3$ to avoid missing local minima due to the chaotic landscape in this problem. We chose a population of 70 with an initial bound of $\pm20$ in their search space, the initial step size was set to $\sigma_0 = 12$, the mean was fixed to 0, and the covariance matrix was the identity matrix. 
\end{itemize}

\end{document}